\documentclass[useAMS,usenatbib,usedcolumn]{mn2e}

\pdfoutput=1
\usepackage{graphicx}
\usepackage{color}
\title[WDM and small-scale inconsistencies]{Warm dark matter does not do better than cold dark matter in solving small-scale inconsistencies}
\author[Schneider et al.]{Aurel Schneider$^{1}$\thanks{Email: aurel.schneider@sussex.ac.uk}, Donnino Anderhalden$^{2}$, Andrea V. Macci\`o$^{3}$, and J\"urg Diemand$^{2}$\\
{$^1$Department of Physics and Astronomy, University of Sussex, Brighton, BN1 9QH, UK}\\
{$^2$Institute for Theoretical Physics, University of Zurich, 8057 Zurich, Switzerland}\\
{$^3$Max Planck Institut f\"ur Astronomie, K\"onigstuhl 17, D-69117 Heidelberg, Germany}}

\begin{document}


\label{firstpage}
\maketitle

\begin{abstract}
Over the last decade, warm dark matter (WDM) has been repeatedly proposed as an alternative scenario to the standard cold dark matter (CDM) one, potentially resolving several disagreements between the CDM model and observations on small scales. Here, we reconsider the most important CDM small-scale discrepancies in the light of recent observational constraints on WDM. As a result, we find that a conventional thermal (or thermal-like) WDM cosmology with a particle mass in agreement with Lyman-$\alpha$ is nearly indistinguishable from CDM on the relevant scales and therefore fails to alleviate any of the small-scale problems. The reason for this failure is that the power spectrum of conventional WDM falls off too rapidly. To maintain WDM as a significantly different alternative to CDM, more evolved production mechanisms leading to multiple dark matter components or a gradually decreasing small-scale power spectrum have to be considered.
\end{abstract}


\section{Introduction}
While the standard model of cosmology, $\Lambda$CDM, is an indisputable success on large scales, it has several potential problems on smaller scales. Examples are the overabundance of dwarf galaxies in the Milky Way galaxy \citep{Klypin1999,Moore1999}, the Local Group \citep{Zavala2009}, and local voids \citep{Tikhonov2009}, or an excess of dark matter in the centres of dwarf galaxies \citep{deBlok2001}. Whether these discrepancies are a result of our poor understanding of galaxy formation or if they are a hint for an alternative form of dark matter is currently under debate \citep{Weinberg2013}.

One of the most popular alternative dark matter scenarios, which seems to naturally solve many of the small-scale disagreements while being indistinguishable from CDM on larger scales, is the warm dark matter (WDM) paradigm, where the power spectrum is characterised by steep cutoff at the dwarf galaxy scales \citep{Bode2001}. Due to the lack of small-scale power, WDM structure formation is suppressed, resulting in a reduced dwarf galaxy abundance and shallower inner profiles, which are in better agreement with observations \citep{Maccio2012b,Schneider2012}.

The most popular candidate for WDM is the sterile neutrino, which naturally arises from a minimal extension of the neutrino sector within the standard model \citep{Merle2013}, motivated by the recently observed neutrino oscillations \citep{Gonzalez-Garcia2008}. In the early universe, sterile neutrinos can be produced via oscillations with active neutrinos \citep{Dodelson1994} which leads to a non-thermal velocity distribution and a transfer function with a characteristic cutoff \citep{Viel2005}. This cutoff has the same shape as the one from thermal production but is slightly shifted towards higher values of $k$, something that can be accounted for by simply restating the effective mass of the WDM particle. It is therefore conventional to give the generic thermal mass of the WDM particle $m_{\rm WDM}$, and to estimate the WDM mass of a specific `thermal-like' production mechanism by comparing the corresponding transfer functions.

In the last decade, a number of different authors have proclaimed WDM as a potential solution to observed discrepancies on small scales, however, with different WDM particle masses depending on the problem.  The excess of dark matter velocity dispersion in the inner parts of Milky Way satellites, the {\it too big to fail} problem \citep{Boylan-Kolchin2012}, has been explained by a WDM model with particle mass of $m_{\rm WDM}=1.4-2$ keV \citep{Lovell2012,Anderhalden2013,Lovell2013}, while the luminosity function from semi-analytical modelling has been shown to be in agreement with a WDM model of $m_{\rm WDM}=0.75$ \citep{Menci2013}. The HI velocity function on the other hand seems to be best matched by a WDM with $m_{\rm WDM}=1.0$ \citep{Papastergis2011}. The use of different WDM particle masses depending on the problem leads to the somewhat misleading impression that WDM provides ideal solutions to different regimes of small-scale structure formation. However, in order to be consistent, a single particle mass needs to provide a solution to all problems and this specific WDM scenario has to pass all observational constraints.

The major difficulty of using small-scale structure formation to constrain the nature of dark matter, is the influence of baryons, which is largely unknown and could produce effects that are degenerate to the expected effects from dark matter \citep{Herpich2013}. For example, photo-evaporation during the epoch of reionization as well as stellar feedback effects from the first stars are expected to blow the gas out of the potential wells of small haloes, to switch off star formation and render them completely dark \citep{Maccio2010}. Alternatively, stellar feedback could also alter the inner dark matter density profiles of dwarf galaxies, it is however questionable whether this effect is large enough to reconcile theory with observations \citep{Garrison-Kimmel2013}.

In this Letter, we reconsider the WDM scenario under the light of the latest constraints from Lyman-$\alpha$ forest and SDSS data. Since these constraints exclude most of the WDM scenarios mentioned above, one should ask the question if the WDM paradigm is still able to alleviate the observed small-scale problems. We directly compare circular velocities from the stellar and HI content of observed dwarf galaxies with the velocity dispersions from $N$-body simulations and analytical models. Since circular velocities trace the underlying dark matter density field, no detailed knowledge of the hydrodynamical interactions is required. The most stringent tests come from the stellar velocities at the half-light radius of dwarf galaxies and from the HI velocities measured with 21cm surveys. We use both of these observations to check whether WDM models within the allowed constraints are still able to solve the small-scale crisis present in CDM structure formation.

\section{WDM particle mass}
The WDM paradigm has been tested with different observations, leading to independent constraints on the WDM particle mass. Examples are the number of dwarf galaxies \citep{Polisensky2010}, weak lensing \citep{Miranda2007,Smith2011}, galaxy formation \citep{Maccio2010b}, the Lyman-$\alpha$ forest \citep{Seljak2006,Viel2006}, or gamma-ray bursts \citep{deSouza2013}. Currently, the most stringent constraints come from the Lyman-$\alpha$ forest with $m_{\rm WDM}>3.3$ keV at the 2$\sigma$ level \citep{Viel2013}. This measurement is based on the recent comparison of high redshift quasar spectra combined with an extended series of hydrodynamical simulations. Another tight constraint comes from the number of ultra-faint dwarf galaxies in the SDSS data, which sets a bound on the WDM mass of $m_{\rm WDM}>2.3$ (2$\sigma$) with a maximum likelihood of $m_{\rm WDM}=4$ keV \citep{Polisensky2010}. Both constraints rule out WDM models with a mass of $m_{\rm WDM} = 1 - 2$ keV, which would be able to resolve some of the CDM small-scale problems.

Based on these considerations, we state that a realistic WDM scenario must have a mass of about $m_{\rm WDM}\sim 4$ keV to be in perfect agreement with both Lyman-$\alpha$ and SDSS data. In the following, we will test whether such a scenario is able to alleviate the CDM small-scale crisis.

\section{Too big to fail}
One of the outstanding problems of CDM structure formation is the deep potential wells of the largest Milky Way satellites, leading to circular velocities that are much larger than the observed half-light velocity dispersions of dwarf galaxies. This is called the {\it too big to fail} problem, referring to the fact that these massive satellites are to `big' in order to `fail' to produce stars due to baryonic feedback effects and hence should be observable. Contrary to the {\it missing satellite} problem, the {\it too big to fail} problem is extremely difficult to solve with hydrodynamical feedback effects \citep{Garrison-Kimmel2013}. However, it might be alleviated by adopting the latest $\Lambda$CDM cosmological parameters \citep{diCintio2011, Polisensky2014} and if the subhalo population around the Milky Way lies in the lower few percent of the halo-to-halo variation  \citep{Purcell2012}. Also, the problem would disappear if the total Milky Way halo mass were considerably smaller than currently expected \citep{Rashkov2012,Wang2012,Purcell2012}, but this seems rather unlikely  \citep{Boylan-Kolchin2013}.

Within a WDM scenario of $m_{\rm WDM}\sim 2$ keV, the {\it too big to fail} problem naturally disappears because of the considerably shallower profiles of the largest WDM dwarf galaxies compared to their CDM counterparts. However, it is unclear whether this is still the case for a realistic WDM particle mass of $m_{\rm WDM}\sim 4$ keV. In order to test this, we perform nested high-resolution $N$-body simulations centred on a Milky Way like halo of $M_{200}\sim1.28\cdot10^{12}$ M$_{\odot}$ (where $M_{200}$ is measured with respect to 200 times the critical density) and analyse the profiles of the largest satellites. For the simulations and the subsequent analysis we use the same setup than in \citet{Anderhalden2013,Anderhalden2013b}, namely the cosmological parameters $\sigma_{8}=0.8$, $n_{\rm s}=0.96$, $\Omega_{\rm m}=0.27$, $\Omega_{\Lambda}=0.73$, and $h=0.7$. The only difference is an improved mass resolution by a factor of 4, corresponding to a simulated particle mass of $3.4\cdot10^4$ M$_{\odot}$, and a gravitational softening of 195 pc. The simulations are performed with {\tt pkdgrav}, a parallel tree-code with multiple moment expansion \citep{Stadel2001} and the halo finding is done with the subhalo finder {\tt 6dfof} \citep{Diemand2006}.

\begin{figure*}
\centering{
   \includegraphics[height=6.2cm]{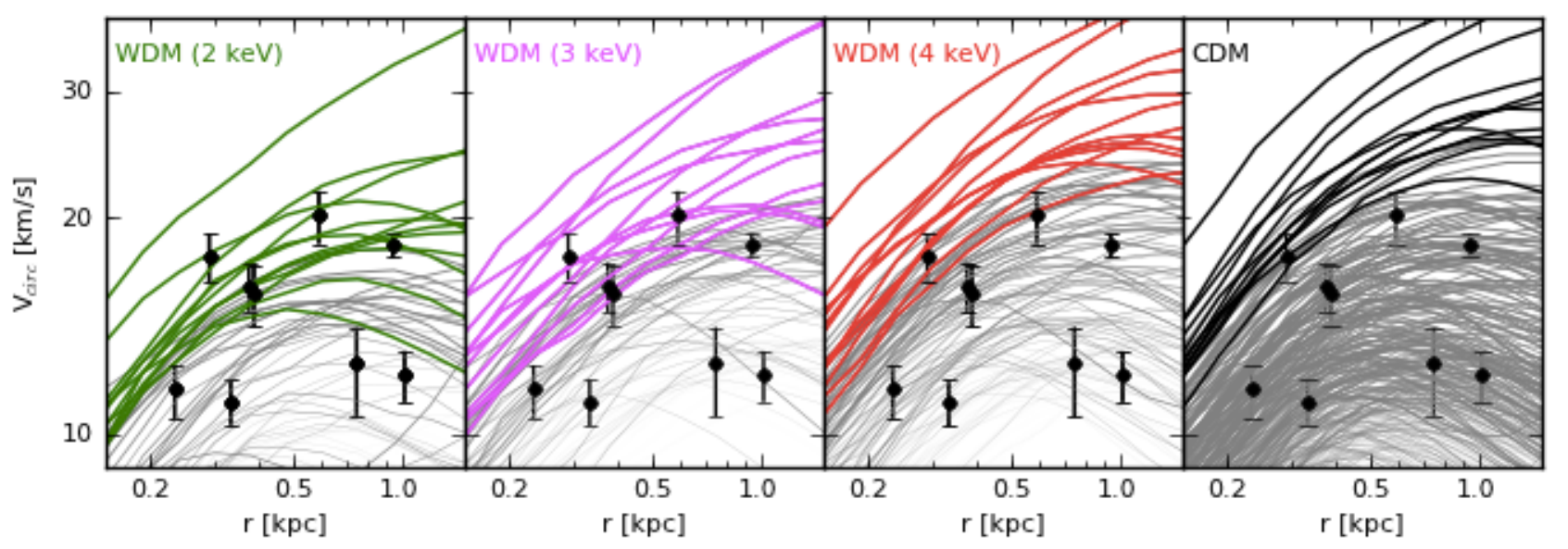}
    }
\caption{\small{Circular velocities profiles of the 12 satellites with the highest $V_{\rm max}$ at infall (green, magenta, red, and black lines). The observed circular velocity at half-light radius of the nine classical dwarfs are added as black dots with error-bars (the LMC, SMC, and Sagittarius are not displayed). From left to right: WDM with $m_{\rm WDM}=2,3,4$, and CDM. The grey lines are the remaining satellites above $V_{\rm max}=12$ km/s with decreasing line width for smaller $V_{\rm max}$ at infall.}}
\label{circvel}
\end{figure*}

In Fig. \ref{circvel}, we plot the velocity profiles of the 12 satellites with the largest $V_{\rm max}$ at infall \citep[see][for more information on the method]{Anderhalden2013b} and compare them to the observed half-light stellar velocity dispersion of nine classical dwarf galaxies (LMC, SMC, and Sagittarius have been removed from the sample). For the CDM case (in the right-hand panel), the profiles (in black) are systematically above the observed dots, showing the standard {\it too big to fail} discrepancy\footnote{Our CDM host-halo contains five satellites with $V_{\rm max}>30$ km/s. This is somewhat fewer than that reported by \citet{Boylan-Kolchin2012} but well within the expected halo-halo scatter, as shown by \citet{Garrison-Kimmel2013b}.}. In the WDM case with $m_{\rm WDM}=2$ keV (plotted in the left-hand panel), the discrepancy disappears and the profiles (in green) roughly coincide with the observed dots. This WDM scenario seems to solve the {\it too big to fail} problem, it is however ruled out by the recent constraints from Lyman-$\alpha$ forest and SDSS data (as discussed in the former section). In the case of a realistic WDM scenario with $m_{\rm WDM}=4$ keV (third panel from the left), the circular velocity profiles (in red) are significantly above the values of the observed dwarf galaxies, yielding a similar picture than in the case of CDM.
From the content of Fig. \ref{circvel} it is clear that a realistic WDM scenario that passes the Lyman-$\alpha$ constraints is too cold to significantly alleviate the {\it too big to fail} problem. During the publication process of this work, \citet{Polisensky2014} released a paper on Milky Way satellites in WDM cosmologies with very similar conclusions. In particular, they found that the profiles of the largest satellites in a 4 keV WDM model are nearly identical to their CDM counterparts.

\section{HI velocity function}
Another test of small-scale structure formation comes from the HI velocity-width function measured in the local universe by recent 21 cm surveys like the Arecibo Legacy Fast ALFA (ALFALFA) survey \citep{Giovanelli2005}. The general shape of the HI velocity-width function is characterised by a power-law decrease followed by an exponential drop-off, whereas the slope of the power-law is significantly shallower than the one expected from CDM structure formation \citep{Trujillo-Gomez2011}. This fact has motivated several authors to consider a shift of the dark matter paradigm and to suggest WDM as a more realistic scenario \citep{Zavala2009,Papastergis2011}. For example, \citet[hereafter Za09]{Zavala2009} used constrained simulations of the local universe to show that a WDM model with $m_{\rm WDM}=1$ keV leads to a velocity function in much better agreement with observations\footnote{Recently, \citet{Obreschkow2013} pointed out that HI surveys could be substantially incomplete due to a very broad dispersion of the HI mass. Using full semi-analytical modelling, they find that the CDM prediction can be brought in agreement with observations for $W_{50}>50$ km/s, which alleviates the {\it flatness} problem of the HI velocity function.}.

In the following, we revisit this finding in the light of the new constraints of the WDM particle mass, in order to test if a realistic WDM model with $m_{\rm WDM}\sim 4$ keV still agrees with observations. Since we are only interested in the general shape of the HI velocity-width function, we will content ourselves with simplified analytical descriptions of the HI content in galaxies, without running expensive hydrodynamical simulations. The essential ingredient of our approach is the WDM halo mass function developed in \citet{Schneider2013}, which is based on the sharp-k window function and works for cosmologies with arbitrary initial power spectra. The functional form is given by
\begin{eqnarray}\label{WDMmassfct}
\frac{dn}{d\log M}&=&\frac{\bar\rho}{M}f(\nu)\frac{1}{12\pi^2\sigma^2(R)}\frac{P_{\rm lin}(1/R)}{R^3},\\
\sigma(R)&=&\int \frac{d\mathbf{k}^3}{(2\pi)^3} P_{\rm lin}(k)\Theta(1-kR),
\end{eqnarray}
where $\Theta$ is the Heaviside step function and $f(\nu)=A\sqrt{2\nu/\pi}(1+\nu^{-p}) {\rm e}^{-\nu/2}$ with $\nu= (1.686/\sigma)^2$, A=0.322, and p=0.3. The halo mass is assigned to the filter scale by the relation $M=4\pi\bar\rho(cR)^3/3$ with $c=2.7$.

From the halo mass function, it is possible to construct the HI velocity-width function using some simplified assumptions. The procedure consists in first constructing the maximum circular velocity function (short: velocity function) of haloes and then connecting the circular velocity to the measured velocity-width of the HI disk.

We construct the halo velocity function in the same way as Za09, an approach initially developed by \citet{Sigad2000}. The recipe is the following: (i) Producing a mock sample of haloes that mimics the halo mass function for WDM cosmologies (given by Eq. \ref{WDMmassfct}).  (ii) Assigning an NFW-profile to each halo with a randomly selected concentration out of a log-normal distribution from  \citet{Maccio2008}. Using the fitting formula from \citet{Schneider2012} to adopt the concentration to the WDM scenario. (iii) Calculating the maximum circular velocity ($V_{\rm max}$) for every mock halo with the help of Eq. 7 in \citet{Sigad2000}. (iv) Binning the haloes with respect to their value of $V_{\rm max}$ in order to obtain $dn/d\log V_{\rm max}$.

\begin{figure*}
\centering{ 
  \includegraphics[width=8.6cm]{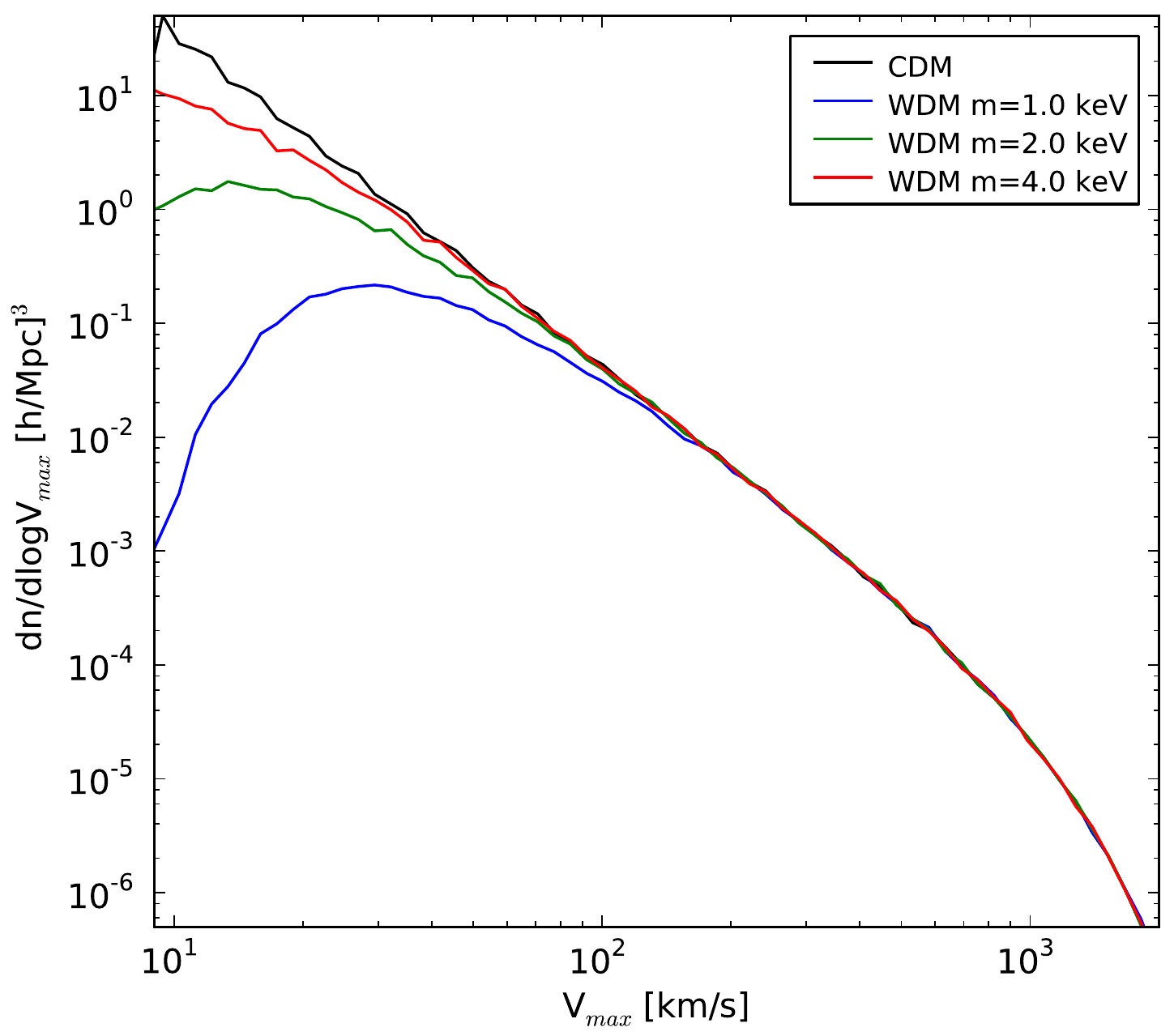}
  \includegraphics[width=8.6cm]{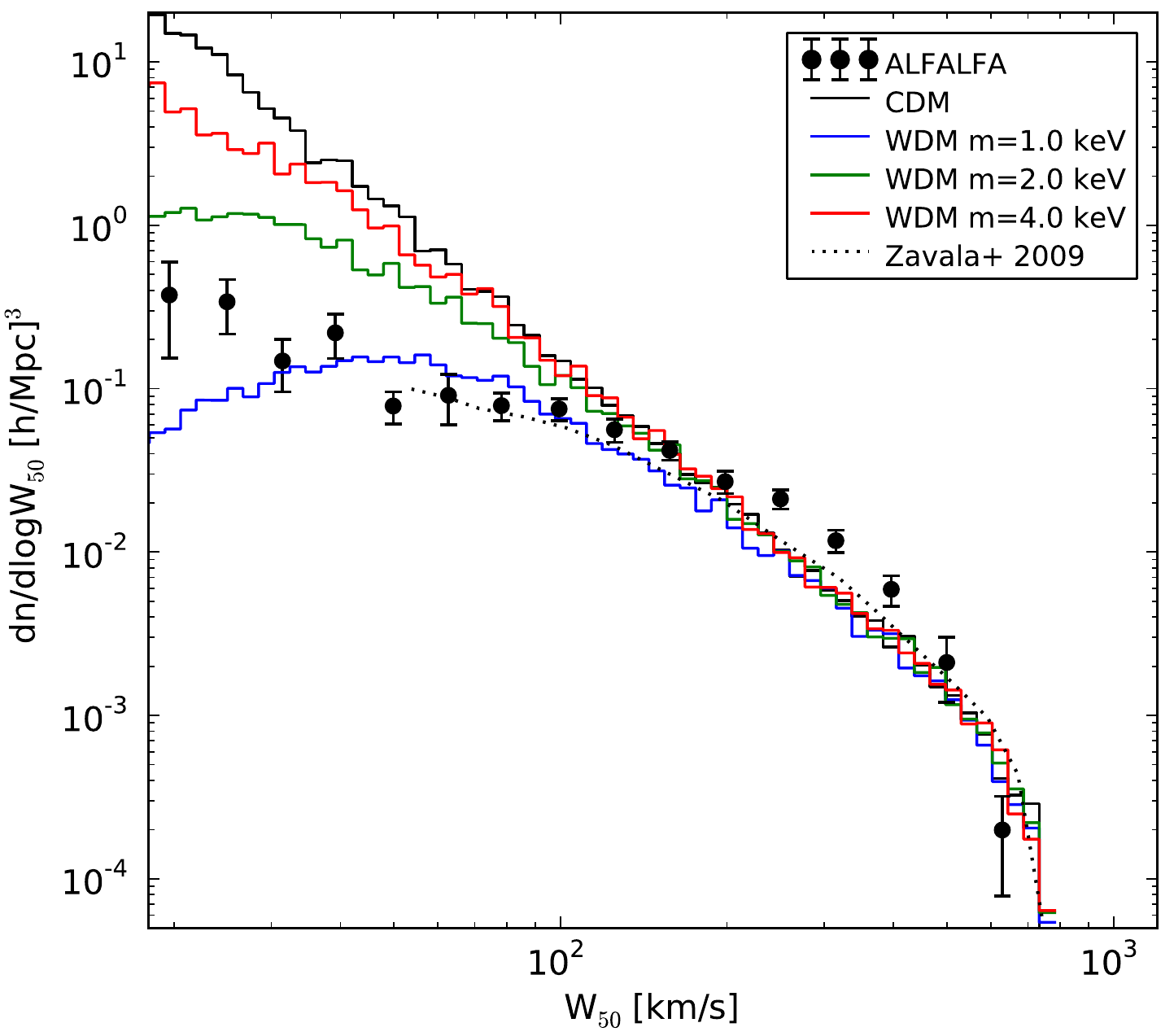}}
\caption{\small{{\it Left-hand panel:} maximum circular velocity function of haloes constructed with the mass function of Eq. (\ref{WDMmassfct}) and with assigned random concentrations from a log-normal distribution. Black: CDM; red: WDM, 4keV; green: WDM, 2keV; blue: WDM 1keV.
{\it Right-hand panel:} Velocity-width function of the HI component measured by ALFALFA (black dots, \citet{Papastergis2011}) and obtained by converting $V_{\rm max}$ into $W_{50}$ as explained in the text. Same colour coding. The simulated velocity-width function from Za09 is given as a black dotted line for comparison.}}\label{velfct}
\end{figure*}

The velocity function of haloes $dn/d\log V_{\rm max}$ is plotted in the left-hand side of Fig. \ref{velfct}, where the blue, green, and red lines represent WDM cosmologies with particle masses of 1, 2, and 4 keV, while the black line represents the standard CDM model. The WDM velocity function has a similar shape to the original halo mass function (plotted in Fig. 5 of \citet{Schneider2013}), with a suppression and a downturn below a certain value of $V_{\rm max}$.

The connection between the halo velocity function and the velocity-width function of the HI component consists in setting HI disks into the mock haloes with an appropriate velocity-width $W_{50}$. We again closely follow Za09, utilising the following recipe: (i) Populating every mock halo below $10^{13}$ M$_{\odot}$/h with exactly one HI disk\footnote{This is motivated by the fact that satellite galaxies have lost their gas due to dynamical friction (see Za09 for a detailed discussion on the validity of this approximation).}. (ii) Calculating the maximum circular velocity of the disk with the help of Eq. (1) from Za09, using a fixed disk-to-halo mass ratio $f=0.03$ and a randomly selected halo spin out of a log-normal distribution from \citet{Maccio2008} for both CDM and WDM simulations\footnote{We have checked the distribution of the spin in WDM simulations presented in \citet{Schneider2012} and found no systematic differences between WDM and CDM haloes.}. (iii) Omitting all haloes with an assigned spin below 0.02 in the sample because no stable disk are expected to form in this regime. (iv) Connecting the velocity-width to the disk circular velocity by setting $W_{50}=2\sin(\rm i)V_{\rm max,d}$, where the disk-inclination is assumed to be a random number in the range $[0,\pi]$. (v) Binning the disks with respect to their value of $W_{50}$ to obtain $dn/d\log W_{50}$.

The velocity-width function is plotted in the right-hand side of Fig. \ref{velfct}, where the blue, green, and red lines represent WDM models with 1, 2, and 4 keV, while the black line represents the CDM cosmology. The observed data from the ALFALFA survey is plotted as black dots. The faint dotted line is the result from Za09, which is based on constrained simulations of the local universe with a 1 keV WDM model and has a resolution limit of $W_{50}=36$ km/s. The right-hand side of Fig. \ref{velfct} shows that the CDM curve is roughly in agreement with observations above $W_{50}=100$ km/s and lies significantly above the observations for smaller velocities. The same is true for the 4 keV and, to a minor extend, for the 2 keV WDM model. The 1 keV WDM model, on the other hand, gives a reasonable match over all velocity scales, as predicted by Za09. At very small scales, $W_{50}<50$ km/s, the predicted WDM velocity-width function turns over, something that is not visible in the data and indicates that the 1 keV model might be too extreme to explain the data\footnote{The slight mismatch between all the models and the observations around $W_{50}=300$ km/s is likely to come from the fact that the ALFALFA data come from an overdense patch of the sky, something that is not accounted for in our model.}. In summary, Fig. \ref{velfct} shows that a realistic WDM model, which passes all constraints from the Lyman-$\alpha$ forest and SDSS data, is not able to provide a significantly better explanation to the apparent flatness of the HI velocity-width function than the standard CDM model.

\section{Conclusions}
We have tested the WDM paradigm on two of the most prominent small-scale problems of CDM structure formation, the  {\it too big to fail} problem of the largest Milky Way satellites and the {\it flatness} problem of the HI velocity-width function in the local universe. As a result, we have shown that a realistic WDM scenario with $m_{\rm WDM}=4$ keV in agreement with recent constraints from Lyman-$\alpha$ forest and SDSS data fails to alleviate the potential small-scale problems of CDM structure formation. The reason for this failure can be attributed to the shape of the cutoff in the linear power spectrum, which is too steep to simultaneously agree with the Lyman-$\alpha$ data and produce a more natural match to the dwarf galaxy observations. Hence, from an astrophysical perspective, there is no convincing reason to favour WDM from thermal or thermal-like production (i.e. neutrinos oscillations) over the standard CDM scenario.

There are however alternative WDM production mechanisms where these conclusions do not necessarily apply. For example, sterile neutrinos could be present as a mixture of non-resonantly produced warmer and resonantly produced colder particles, leading to a shallower downturn in the power spectrum \citep{Boyarsky2009,Boyarsky2009b}. Recent studies on the structure formation of such models seem promising \citep{Maccio2012,Anderhalden2013,Anderhalden2013b, Marsh2013}, but more investigation is necessary to test whether these alternative approaches agree with both Lyman-$\alpha$ forest and ultra faint dwarf galaxies.

In general, a deeper understanding of galaxy formation is required to obtain a more conclusive view on the currently stated small-scale discrepancies. Including self-consistent hydrodynamics will be crucial to further constrain the nature of dark matter with astrophysical observations.

We thank Darren Reed and William Watson for useful discussions. AS acknowledges support from the European Commissions Framework, through the Marie Curie Initial Training Network CosmoComp (PITNGA-2009-238356). All simulations have been performed on the SuperMuc and Theo clusters in Munich and the zbox4 cluster in Zurich.

\end{document}